\documentclass{article}
\usepackage{amsmath,amssymb}
\usepackage{cite,a4wide}

\DeclareMathOperator{\tr}{\mathrm{Tr}}

\title{Nodal Liquids and Duality}
\author{Nick E. Mavromatos and Sarben
Sarkar\\{\hspace{1cm}}\\{\textsl{Department of Physics, King's College
London,}}\\{\textsl{Strand, London WC2R 2LS, United Kingdom.}}}
\date{ }

\begin{document}

\maketitle

\begin{abstract} 
Using a SU(2) $\times $ U(1) gauge theory for a t--J model around a
node of the Fermi surface, we discuss patterns of dynamical symmetry
breaking, which may lead to a pseudogap phase and to the appearance of
narrow one-dimensional spatial structures, induced by the presence of
holes. A possible connection with stripe phases is briefly discussed
by passing to an appropriate dual theory. We discuss confinement
properties of spinons and holons and derive the spectrum of the
(gapful) physical excitations which are composites of holons
corresponding to `mesons', and composites of spinons corresponding to
`baryons'.
\end{abstract} 

\vspace*{-4in}
\begin{flushright}
cond-mat/0006234
\end{flushright}
\vspace*{4in}

Strong electron correlations are regarded generally as an important
ingredient of a theory for high temperature superconductivity.  There
has been a great deal of ingenuity in theoretical approaches and this
has been concentrated on the most appropriate way of incorporating
constraints in terms of `particle' occupancy at sites of the
underlying lattice.  Symmetry, both global and local and both in the
Wigner--Weyl mode or the Nambu--Goldstone mode have been shown to be
essential ingredients.  However the exact symmety group and the
representation and nature of any symmetry breaking forms are the
questions which need to be settled.

In this note we will put forward a candidate for this symmetry~\cite{farakos}:
\begin{gather} \label{gsymm}
G\simeq \mathrm{SU(2)}\times\mathrm{U(1)},
\end{gather}
where $G$ is a local symmetry.  
The SU(2) symmetry is present in a
Nambu--Goldstone form and only a compact U$_\tau$(1) associated with
the $\tau_3$ generator of SU(2) remains unbroken. In the physical 
(2+1)-dimensional 
model~\cite{farakos} the statistical gauge symmetry U(1) appears
in a gauge fixed form, and in fact is associated with exotic statistics
of the relevant excitations. 
This basic symmetry breaking which is achieved dynamically~\cite{farakos} 
leads to a gauge-invariant 
mass gap for charge degrees of freedom (the pseudogap) and also to
`one dimensional' structures which we suggest are related to the
insulating stripe phases which are of current interest in underdoped
cuprates.  In addition there is a global SU(2) symmetry related to the
spin which in the gauge theory approach will be associated with a
`custodial' SU(2) global symmetry, as we shall discuss at the end.

Our aim is, starting from a reasonable microscopic Hamiltonian, $H$,
to form a continuum field theory for the low energy excitations of the
cuprates.  Such a model belongs to a class of t--J models with
additional interactions~\cite{farakos,sarben}.  
At half-filling $H$ reduces to the
Heisenberg Hamiltonian for which adding a spin-up particle is the same
as removing a spin-down particle owing to the constraint of single
particle occupancy.  This in turn leads to a `hidden' SU(2) gauge
symmetry of the Heisenberg model.  In the presence of holes this no
longer holds, and the theory has a U(1) gauge symmetry.  Usually there
is a `discontinuity' in the formulation of the problem at half-filling
and away from half-filling.  In the former there is a local SU(2)
symmetry while in the latter the SU(2) formulation is totally replaced
with a U(1) local symmetry from the outset.  In our case we have the
symmetry group $G$ which reduces to SU(2) at half-filling while away
from half-filling, owing to strong gauge fluctuations associated with
the U(1) factor in $G$, the SU(2) is dynamically broken down to
U$_\tau$(1), i.e. the system decides about the local symmetry given a
maximal $G$-symmetry.

In the two-dimensional Heisenberg model 
Affleck \textsl{et al}~\cite{affleck} noted
that
\begin{gather}
\underline{S}_i = \tr \left( \chi_i^\dagger \chi_i
\;\underline{\sigma}^\mathrm{T} \right)
\end{gather}
where, in terms of $c_1$ and $c_2$ (the annihilation operators for up
and down electrons),
\begin{gather}
\chi = \begin{pmatrix} c_1 & \phantom{-}c_2 \\ c_2^\dagger &
-c_1^\dagger \end{pmatrix}
\end{gather}
and so there is a local SU(2) symmetry
$\chi_i \longrightarrow h_i \chi_i,~~ h_i \in \mathrm{SU(2)}$.
In all the above the index $i$ labels the lattice sites and
$\underline{\sigma}$ is the vector of Pauli-matrices.

Away from half-filling we make the \textsl{ansatz}~\cite{farakos}
\begin{gather}
\chi = \begin{pmatrix} \psi_1 & \phantom{-}\psi_2 \\ \psi_2^\dagger &
-\psi_1^\dagger \end{pmatrix} \begin{pmatrix} z_1 & -\overline{z}_2 \\
z_2 & \phantom{-}\overline{z}_1 \end{pmatrix}
\label{ansatz} 
\end{gather}
where $\psi_\alpha$, $z_\alpha$ with $\alpha\in\{1,2\}$ are fermions
and bosons respectively.  The $\psi_\alpha$ describe charge degrees of
freedom and the $z_\alpha$ describe the spin degrees of freedom.  The
index $\alpha$ is related to the underlying bipartite
(antiferromagnetic) lattice structure.  At a site the original
formulation had two degrees of freedom: the spin-up hole and the
spin-down hole.  The above spin-charge separation \textsl{ansatz}
should have constraints and symmetries which reduce the number of
physical degrees of freedom to two.  One constraint is that $c_\alpha$
should obey canonical commutation relations.  On assuming canonical
commutation/anti-commutation relations for $z_\alpha$ and
$\psi_\alpha$ we find that the canonical anticommutation relations for
the $c$ operators are satisfied on physical states provided the
following constraints at each site $i$ hold~\cite{farakos}:
\begin{gather}
\psi_{1,i} \; \psi_{2,i} =0= \psi_{2,i}^\dagger \;
\psi_{1,i}^\dagger~,~
 \sum_{\beta=1,2} \left( \overline{z}_{i,\beta} \;z_{i,\beta} +
\psi_{\beta,i}^\dagger \; \psi_{\beta,i} \right) =1.
\end{gather}
These relations are single occupancy constraints and imply that our
formulation is suitable in the limit of strong correlations.  It will
be convenient to work in the functional integral framework where the
$\psi$ are Grassmann variables.  If we count the degrees of freedom in
$\chi$, $\deg \chi$, we have
$\deg \chi = 4 + 4$, 
where 4 counts the Grassmann degrees of freedom and 4 is the number of
real degrees of freedom in $z_1$, $z_2$, $\overline{z}_1$ and
$\overline{z}_2$.

Just as in the Heisenberg case there is an SU(2) symmetry
\begin{align}
\Psi_i &\longrightarrow \Psi_i \; h_i \qquad \text{where} \qquad \Psi =
\begin{pmatrix} \psi_1 & \phantom{-}\psi_2 \\ \psi_2^\dagger &
-\psi_1^\dagger \end{pmatrix} \notag \\
Z_i &\longrightarrow h_i^\dagger \; Z_i \qquad \text{where} \qquad Z =
\begin{pmatrix} z_1 & -\overline{z}_2 \\ z_2 &
\phantom{-}\overline{z}_1 \end{pmatrix}
\end{align}
where $h_i \in$ SU(2). However, there is also a dynamical U(1) gauge
symmetry acting on the $\Psi$ fields, which is due to phase
frustration from holes moving in a spin background.  Some arguments to
justify this from a microscopic point of view have been given in
reference~\cite{weng+dorey}.  Consequently, this symmetry is
associated with exotic statistics of the pertinent
excitations~\cite{farakos}, which is an exclusive feature of the
planar spatial geometry.  The existence of these gauge symmetries
together with the constraints reduce the effective number of degrees
of freedom to the physical degrees of freedom of the system.  The
\textsl{ansatz} gives us a maximal symmetry $G$.  The dynamics will be
specified by a Hamiltonian which will determine any spontaneous
symmetry breaking that occurs.

As a generic model we will consider a generalized t--J model which is
an effective single-bond model derived from a more realistic
5-bond model~\cite{sarben}.  
This is in terms of the parameters $\{t,t',t'',J,V\}$
where $t$, $t'$, and $t''$ are nearest-neighbour, next
nearest-neighbour and third nearest-neighbour hoppings respectively,
$J$ is the Heisenberg antiferromagnetic interaction and $V$ is a
static attractive nearest-neighbour interaction.  It turns out that,
under certain circumstances, the presence of $V$ allows the theory in
the continuum limit to show dynamical supersymmetry 
between spinon and holon degrees of freedom as has been
discussed in ref. \cite{sarben}.

The Heisenberg term can be written 
in terms of a Hubbard--Stratonovich field,
$\Delta_{ij}$, as
\begin{gather}
-\frac{J}{8} \sum_{\langle ij\rangle} \tr \left( \chi_i \,
 \chi_j^\dagger \, \chi_j \, \chi_i^\dagger \right)=
\sum_{\langle ij\rangle} \tr \left[ \frac{8}{J} \Delta_{ij}^\dagger \,
\Delta_{ji} + \left( \chi_i^\dagger \, \Delta_{ij} \, \chi_j +
\mathit{h.c.}\right) \right].
\end{gather}
The hopping part of the Hamiltonian can be written as
\begin{multline}
-\sum_{\langle ij\rangle} t_{ij} \, c^\dagger_{\alpha,i} \;
 c_{\alpha,j} =
-\sum_{\langle ij\rangle} t_{ij} \left( \chi^\dagger_{i,\alpha\gamma}
 \; \chi_{j,\gamma\alpha} + \chi^\dagger_{i,\alpha\gamma} \,
 (\sigma_3)_{\gamma\beta}\, \chi_{j,\beta\alpha} \right)=\\
= -\sum_{\langle ij\rangle} t_{ij} \left( \overline{Z}_{i,\beta\kappa}
 \, \Psi^\dagger_{i,\kappa\alpha} \, \Psi_{j,\alpha\gamma} \,
 Z_{j,\gamma\beta} + \overline{Z}_{i,\beta\kappa}
 \, \Psi^\dagger_{i,\kappa\alpha} \,(\sigma_3)_{\alpha\lambda} \Psi_{j,\lambda\gamma} \,
 Z_{j,\gamma\beta} \right).
\end{multline}
The global SU(2) spin symmetry, mentioned above, is explicitly given
by \( Z_i \rightarrow Z_i h\) (and equivalently \(\chi_i \rightarrow
\chi_i h\)) where \(h\in\) SU(2) is a group element.

In the Hartree--Fock approximation we obtain the Hamiltonian
\begin{multline}
H_\mathrm{HF} = \sum_{\langle ij\rangle} \tr \left[ \frac{8}{J}
\Delta^\dagger{ij} \, \Delta_{ji} + \left( -t_{ij} (1+\sigma_3) +
\Delta_ij\right) \Psi_j^\dagger \langle Z_j \, \overline{Z}_i\rangle
\Psi_i\right] +\\
+ \sum_{\langle ij\rangle} \tr \left[ \overline{Z}_i \langle
\Psi^\dagger_i \left( -t_{ij} (1+\sigma_3) + \Delta_ij\right)
\Psi_j\rangle Z_j + \mathit{h.c.}\right].
\end{multline}

Using the $G$-symmetry~\eqref{gsymm} of the \textsl{ansatz} we can
write in the presence of gauge fixing 
\begin{gather}
\langle Z_j \overline{Z}_i \rangle \equiv |A_1| \, {\cal R}_{ij} \, U_{ij}~,~
\langle \Psi^\dagger_i \left( -t_{ij} (1+\sigma_3) +\Delta_ij\right)
\Psi_j \rangle \equiv |A_2| \, {\cal R}_{ij} \, U_{ij}.
\end{gather}
where ${\cal R}\in$ SU(2) and $U\in$ U(1) are group elements. The fact
that apparently gauge non-invariant correlators are non-zero on the
lattice is standard in gauge theories, and does not violate
Elitzur's theorem~\cite{elitzur}, precisely due to the above-mentioned
gauge-fixing procedure, which is done prior to any computation.  The
amplitudes $|A_1|$ and $|A_2|$ are considered frozen which is a
standard assumption in the gauge theory approach to strongly
correlated electron systems~\cite{affleck}.

By standard arguments the low energy lattice action for the fermion
part becomes
\begin{gather}
S_\mathrm{F} = \frac{1}{2} \kappa' \sum_{i,\mu} \left(
\overline{\hat{\Psi}}_i \,\gamma_\mu \, {\cal R}_{i,\mu}\,U_{i,\mu}\,
\hat{\Psi}_{i+\mu} + \overline{\hat{\Psi}}_{i+\mu} \,\gamma_\mu \,
{\cal R}_{i,\mu}\,U_{i,\mu}\, \hat{\Psi}_i \right)
\end{gather}
where $\hat{\Psi}_{i,\alpha}$ is a Nambu--Dirac two-component spinor
for each `colour' $\alpha\in\{1,2\}$.  The ${\cal R}$ and $U$ matrices
act on the colour matrices whereas the $2\times 2$ $\gamma$ matrices
act on the two components of $\hat{\Psi}_{i,\alpha}$.  The components
of $\hat{\Psi}$ are linearly related to those of $\Psi$ Moreover, we
have kinetic (i.e. Maxwell) terms for the gauge link variables in the
form of plaquette terms in the lattice action:
\begin{gather}
S_\mathrm{G} = \sum_p \left[ \beta_\mathrm{2} (1-\tr {\cal R}_p) +
\beta_\mathrm{1} (1-\tr U_p) \right]
\label{plaquette}
\end{gather}
where $p$ denotes plaquettes, the $\beta_i$ are inverse couplings,
$\beta_2\equiv \beta_\mathrm{SU(2)}\propto 1/g^2$, $\beta_1 \equiv
\beta_\mathrm{U(1)}$, and ${\cal R}_p$, $U_p$ are a product of the
link variables over the plaquette $p$.  It is important to notice for
our purposes that, as a result of the gauge-fixed form of the U(1)
factor in $G$ in the spinon sector, there is no plaquete term for the
U(1) field, which implies $\beta_1=0$ and hence we are in the strong
coupling limit~\cite{farakos}.  The coupling $\beta_2$ is
large~\cite{sarben}.

We will find it useful to give the explicit representation of the
global $G$-symmetry whose gauging produces $S_\mathrm{G}$.  On writing
\begin{gather}
\tilde{\Psi} = \begin{pmatrix} \hat{\Psi}_1 \\ \hat{\Psi}_2
\end{pmatrix}
\end{gather}
the generators of $G$ are given by
\begin{gather}
\left\{ \boldsymbol{1}_4, \begin{pmatrix} 0& \boldsymbol{1}_2 \\
\boldsymbol{1}_2 & 0 	  \end{pmatrix} , i\begin{pmatrix} \phantom{-}0 &
\boldsymbol{1}_2 \\ -\boldsymbol{1}_2 & 0  \end{pmatrix},
\begin{pmatrix} \boldsymbol{1}_2 & \phantom{-}0 \\ 0 &
-\boldsymbol{1}_2\end{pmatrix} \right\}.
\end{gather}
Let us consider some relevant multiplets of this group at any site
$j$.  The bilinears
\begin{equation} 
\phi_1 = -i ( \overline{\hat{\Psi}}_1 \hat{\Psi}_2 -
\overline{\hat{\Psi}}_2 \hat{\Psi}_1 )~,~
\phi_2 = \overline{\hat{\Psi}}_1 \hat{\Psi}_2 +
\overline{\hat{\Psi}}_2 \hat{\Psi}_1~,~
\phi_3 = \overline{\hat{\Psi}}_1 \hat{\Psi}_1 -
\overline{\hat{\Psi}}_2 \hat{\Psi}_2
\end{equation}
form an adjoint representation of SU(2)~\cite{farakos}.  
There is also a vector
adjoint representation 
\begin{align}
({\cal A}_\mu)_1 &= i ( \overline{\hat{\Psi}}_1 \,\tilde{\sigma_\mu}\,
\hat{\Psi}_2 - \overline{\hat{\Psi}}_2
\,\tilde{\sigma_\mu}\,\hat{\Psi}_1 )~,~
({\cal A}_\mu)_2~&= \overline{\hat{\Psi}}_1 \,\tilde{\sigma_\mu}\,\hat{\Psi}_2 +
\overline{\hat{\Psi}}_2 \,\tilde{\sigma_\mu}\,\hat{\Psi}_1 , \notag\\
({\cal A}_\mu)_3 & =\overline{\hat{\Psi}}_1 \,\tilde{\sigma_\mu}\,\hat{\Psi}_1 -
\overline{\hat{\Psi}}_2 \,\tilde{\sigma_\mu}\,\hat{\Psi}_2
\end{align}
and two singlets
\begin{equation} 
{\cal S}_4 = \overline{\hat{\Psi}}_1 \hat{\Psi}_1 +
\overline{\hat{\Psi}}_2 \hat{\Psi}_2~,~
({\cal S}_\mu)_4 = \overline{\hat{\Psi}}_1 \,\tilde{\sigma_\mu}\,\hat{\Psi}_1 +
\overline{\hat{\Psi}}_2 \,\tilde{\sigma_\mu}\,\hat{\Psi}_2
\end{equation}
where $\tilde{\sigma_0}= -i\sigma_3$, $\tilde{\sigma_1}=\sigma_1$ and
$\tilde{\sigma_2}=\sigma_2$.

Now we will consider spontaneous symmetry breaking in $G$ due to the
interactions.  
Since
we are at the strong coupling limit of U(1), where
$\beta_1=0$,  the U(1) gauge field can be integrated out
exactly to give~\cite{farakos}
\begin{gather*}
\int \frak{D} {\cal R} \, \frak{D} \overline{\hat{\Psi}}\, \frak{D}
\hat{\Psi} \; \exp (-S_\mathrm{eff})
\end{gather*}
where
\begin{align}
S_\mathrm{eff} &= \beta_2 \sum_p (1-\tr {\cal R}_p) + \sum_{i,\mu} \ln I_0
(\sqrt{y_{i\mu}}),\\
y_{i\mu} &= -\kappa^2 \tr \left( M^{(i)} \,(-\gamma_\mu) \,{\cal R}_{i\mu}
\,M^{(i+\mu)} \,\gamma_\mu \, {\cal R}^\dagger_{i\mu}\right)\\
\intertext{and}
M^{(i)} &= \sum_{a=1}^3 \phi_a(i) \sigma_a + {\cal S}_4(i)
\boldsymbol{1} + i\left( ({\cal S}_\mu)_4(i) \gamma^\mu + \sum_{a=1}^{3}
({\cal A}_\mu)_a(i) \gamma^\mu \sigma_a \right).
\end{align}
Owing to the Grassmann content of $M$,
\begin{gather}
-\ln I_0^{\mathrm{tr}} (2\sqrt{y_{i\mu}}) = -y_{i\mu} + \frac{1}{4}
y_{i\mu}^2 - \frac{1}{9} y_{i\mu}^3 + \frac{11}{192} y_{i\mu}^4,
\end{gather}
which is exact.  This type of analysis leads to
\begin{align}
\langle M^{(i)}\rangle &= u \sigma_3\\
\intertext{and}
\ln I_0^{\mathrm{tr}} (2\sqrt{y_{i\mu}}) & \sim M^2_\mathrm{B} \left(
(B^1_{i\mu})^2 +(B^2_{i\mu})^2 \right) + \text{interaction terms}
\intertext{with $M^2_\mathrm{B} = \kappa^2 u^2$, and so}
\langle \phi_3 \rangle &\neq 0.
\end{align}
(Here \( {\cal R}_{i\mu} = \cos (|\underline{B}_{i\mu}|) +
i\underline{\sigma}\cdot\underline{B}_{i\mu} \sin (
|\underline{B}_{i\mu}|) / |\underline{B}_{i\mu}| \)). 
Notice that since $u \propto \beta_1^{-1}$, in the strong coupling 
limit $u \rightarrow \infty$.  
Hence, a parity
invariant mass for the fermions is generated
dynamically by the strongly-coupled U(1) interactions 
in $G$~\cite{farakos},
and thus the theory is gapped to charge excitations.
But two of the SU(2) gauge bosons
also acquire masses, 
and hence the SU(2) group is broken down to
U$_\tau$(1).  

To make analytical progress towards an understanding of the
confinement properties of spinon and holon degrees of freedom, it is
convenient to go to a continuum limit for the low-energy excitations
of the theory.  By linearizing about a node of the Fermi surface on
the underlying theory of fermions we can consider a relativistic field
theory of the form
\begin{equation}
\frak{L} = -\frac{1}{4} F^a_{\mu\nu} F^{a\,\mu\nu} + \frac{1}{2}
(D_\mu^{ab} \phi^b)^2 + \mu^2 \phi^2 - \lambda (\phi^2)^2
\label{adjointhiggs}
\end{equation} 
where $D^{ab}_\mu \phi^b = \partial_\mu \phi^a -ig
\epsilon^{abc}B^b_\mu \phi^c$ and $ \mu^2 > 0 $, since we have
symmetry breaking (actually, in (2+1)-dimensions the symmetry broken
phase is connected analytically to the phase where the symmetry is
unbroken~\cite{kovner}). The above form should not be considered as
quantitative, but it captures correctly the qualitative features of
our approach, and it is sufficient for our purposes here. Terms
involving ${\cal A}$ and ${\cal S}$ are not important for our
analysis.

At the perturbative level $B_{3\mu}$ is massless, and in fact the
theory is superconducting~\cite{farakos}.  However this is not true
when non-perturbative effects are taken into account, such as
monopoles, which are instantons in the (2+1)-dimensional
theory~\cite{polyakov}.  The monopole is a Euclidean configuration
which behaves asymptotically as
\begin{gather}
\hat{\phi}^a = \hat{r}^a, \qquad \tilde{F}^a_\mu (x) = \frac{1}{g}
\frac{\hat{r}^a \hat{r}_\mu}{r^2}
\end{gather}
where the caret indicates a unit vector and the tilde indicates the
dual field tensor.

At the perturbative level the current $J^\mu$ associated with
U$_\tau$(1) is conserved and in general is given by
\begin{gather}
J^\mu = \frac{1}{g} \epsilon^{\mu\nu\lambda} \partial_\nu (
\tilde{F}^a_\lambda \, \hat{\phi}^a ).
\end{gather}
In the context of the 
model (\ref{adjointhiggs}), it has been shown~\cite{kovner} that 
there exists a local gauge invariant field
$V(x)$ such that
\begin{gather}
J^\mu = -\frac{i}{4\pi} \epsilon^{\mu\nu\lambda} \partial_\mu ( V^*
\partial_\lambda V - \mathit{c.c.}).
\end{gather}
The field $V$ interpolates between defect structures and is a disorder
variable.  In particular we can identify
\begin{gather}\label{Videntification}
 V^* \partial_\lambda V - \mathit{c.c.} = \frac{4\pi i}{g}
 \tilde{F}^a_\lambda \hat{\phi}^a.
\end{gather}
At the perturbative level the current $\tilde{F}_\mu$ is given by
\begin{gather}
\tilde{F}_\mu = \tilde{F}_\mu \hat{\phi}^a - \frac{1}{g}
\epsilon^{\mu\nu\lambda} \epsilon^{abc} \hat{\phi}^a \, (D_\nu
\hat{\phi})^b \, (D_\lambda \hat{\phi})^c
\end{gather}
and is conserved.  However in the presence of monopoles
\begin{gather}
\partial^\mu \tilde{F}_\mu = \frac{4\pi}{g} \delta^{(3)}(x)
\end{gather}
and conservation is lost; the magnetic flux
\begin{gather}
\Phi = \int d^3x\; \tilde{F}_0(x)
\end{gather}
does not generate symmetries of the system, i.e.
$U_\alpha = \exp(i\alpha \Phi)$
is not a symmetry of the physical Hilbert space 
for arbitrary $\alpha$.  Since only
configurations with an integer number of monopoles have finite energy
the discrete flux transformations
\begin{gather}
U_k = \exp(\frac{ig}{2} k \Phi)
\end{gather}
where $k$ is an integer are still symmetries~\cite{kovner}.  
Since the fundamental
flux is in units of $2\pi/g$ the only independent operators on the
Hilbert space are $U_0=\boldsymbol{1}$ and $U_1$:
\begin{equation}
U_0 \, V(x) \, U_0^{-1} = V(x) \qquad
U_1 \, V(x) \, U_1^{-1} = -V(x)
\end{equation}
and so the theory retains a $\mathbb{Z}_2$ symmetry.  The effective
Lagrangian for $V$ is~\cite{kovner}
\begin{gather}
\frak{L}^\mathrm{dual} = \partial_\mu V^* \, \partial^\mu V - \lambda
( V^*\! V \!-\! \mu^2)^2 - \frac{m^2}{4} (V^2\!\!+\! {V^*}^2 ) +h
(\epsilon^{\mu\nu\lambda} \partial_\nu V^* \, \partial_\lambda V)^2.
\end{gather}

The parameters appearing in $\frak{L}^\mathrm{dual}$ can be calculated
perturbatively as
\begin{eqnarray}
&~& \mu^2 = \frac{g^2}{8\pi^2}~,~
\lambda = \frac{2\pi^2 M_\phi^2}{e^2}~,
~h\propto \frac{M_\mathrm{B}}{g^4 M^2_\phi}~,\notag\\
&~& m = m_\mathrm{ph} \propto e^{-S_0/2} \propto M^2_\mathrm{B} 
\exp (-M_\mathrm{B}/g^2)~,
\end{eqnarray}
where $S_0$ is the instanton action.  At low energies the derivative
term can be ignored. The mass $M_\phi$ is that associated with the
$\phi$ fields from the Higgs mechanism; in our 
strong-coupling U(1) situation $M_{\phi} >> m_{ph}$~\cite{farakos}. 
Notice that the presence of a small `photon' mass $m_{ph}$ 
implies a pseudogap phase for the statistical model. 
Because of the
$\mathbb{Z}_2$ symmetry, when external (adjoint) charges are introduced into
the theory, narrow string-like structures are 
produced~\cite{kovner}.  
It is tempting to conjecture that such structures are related
to stripe phases in the cuprates. The width of the stripes
in our model is of order~\cite{kovner} $1/m_{ph}$, which is
finite in underdoped situations. 
The mass 
$m_{ph} \rightarrow0$ as superconductivity is approached, and the stripes
will become very wide and eventually 
occupy the entire space (`absence of stripes').  
Because in equation~\eqref{Videntification}
there is the unit vector 
$\hat{\phi}^a$, which is related to a bilinear in the holon fields, 
the
string-like object requires holons.  Away from the strip region the
Wilson loop shows an area law which is consistent with
antiferromagnetic order. This is related to the fact that,
as we shall discuss below, the spinon excitations are in the fundamental 
representation of the SU(2) group, and as such they are 
directly associated with the Wilson loop which gives 
the energy of a state with 
two heavy external charges in the fundamental representation. 
A detailed correspondence 
requires further research. It should be stressed, though, that 
the above dual description provides a natural and clear 
picture for the 
confinement of both holons and spinons. For details we refer the reader to 
the discussion in ref. \cite{kovner}, which parallels our case here. 

We would now like to discuss the effects and (confinement) properties
of spinons, $z_\alpha$, $\alpha\in\{1,2\}$. From the construction of
the ansatz (\ref{ansatz}) it becomes clear that the spinon sector in
our theory consists of a complex scalar doublet, in the fundamental
representation of the SU(2) gauge group. In the terminology of
ref. \cite{kovner}, this situtation corresponds to including scalar
constituent `quarks' $z^A $ , where $A\in\{1,2\}$ is an SU(2)
fundamental representation index.  The presence of such constituents
leads to the presence of `baryons' in our picture. In other words, in
our model holon composites correspond to `mesons', whilst spinon
composites correspond to `baryons'. These are the physical excitations
of our spin-charge separated non-Abelian gauge model for the planar
doped antiferromagnet.  In the dual picture, the presence of `baryons'
is described~\cite{kovner} by the introduction of an \emph{additional}
field $W$, which should be such that~:~(i) the elementary defect
(soliton) of the field $V$ should carry `baryon' number 1/2 (in the
SU(2) theory) in order to represent the ``constituent (fundamental)
quark''.  For this it is necessary that in the core of the defect
(where $V$ vanishes) the field $W$ has a non-zero value, whilst
outside the core $W \rightarrow \langle W\rangle=0$, since `baryon'
number symmetry cannot be broken spontaneously. (ii) The interaction
potential should favour configurations in which, for $V$ close to its
vacuum expectation value, $W$ is small, whereas for small $V$, $W$ is
non zero. This can be easily implemented by either imposing the
$\sigma$-model constraint $V^*V + W^*W = 1/8\pi^2\beta_2$, or adding
the interaction term $\lambda (V^*V + W^*W - 1/8\pi^2\beta_2)^2$ in
the Lagrangian.  For convenience we choose the $\sigma$-model
constraint.  The long-distance properties, we are interested in, are
indistinguishable between the two cases~\cite{kovner}.  (iii) The
`baryon' should be attached to the topological soliton, so
three-derivative terms are necessary in the dual effective Lagrangian,
which on symmetry grounds has the form~\cite{kovner}:
\begin{eqnarray} 
&&\frak{L}^\mathrm{dual/baryon} = \partial_\mu V^* \, \partial^\mu V 
+ \partial_\mu W^* \, \partial^\mu W +h (V + V^* ) + \nonumber \\
&&\qquad\qquad\qquad\qquad +\frac{1}{4W^*W}(W^*\partial_\mu W -
W\partial_\mu W^*)^2 -\\
&&\:-\frac{1}{8\pi (W^*W)(V^*V)}\epsilon^{\mu\nu\lambda} (W^*\partial_\mu
W - W \partial_\mu W^*)~ \partial_\nu (V^*\partial_\lambda V - V
\partial_\lambda V^*) \nonumber
\end{eqnarray}
We now remark that it is an exclusive feature of the SU(2) case that
the vacuum in the one-soliton sector is degenerate~\cite{kovner},
corresponding to solitons with `baryon' number $1/2$ and $-1/2$.  This
degeneracy corresponds to an additional global SU(2) `custodial'
symmetry~\cite{kovner}, under which the $Z$-matrix in the ansatz
(\ref{ansatz}) is transformed as $Z \rightarrow Z U$.  The `baryon'
number is therefore associated with one of the generators of the
custodial SU(2). The fermion matrix is neutral under the custodial
symmetry.  In the condensed-matter framework this symmetry may be
associated with the global SU(2) spin group of the electron
constituents mentioned earlier.  Evidently, `particles' in this dual
picture come in pairs with opposite `baryon' number.  In the dual
picture the confinement of spinons is evident~\cite{kovner}, as can be
demonstrated by means of the area law behaviour of the Wilson loop.
This completes our qualitative discussion on the symmetry-breaking and
confinement properties of our (zero- (or low-) temperature) effective
spin-charge separated theory. More work is clearly needed for a
quantitative analysis.

\subsection*{Acknowledgements}

We thank A.~Kovner and A.~Campbell--Smith for discussions.  This work
is supported in part by the Leverhulme Trust; N.E.M. is also partially
supported by P.P.A.R.C. (U.K.).

\end{document}